\documentclass[12pt]{article}

\newcommand{\br}{\hskip .25cm/\hskip -.25cm}
\begin{document}

\sloppy
\begin{flushright}{SIT-HEP/TM-7}
\end{flushright}
\vskip 1.5 truecm
\centerline{\large{\bf Activated Sphalerons and Large Extra Dimensions}}

\vskip .75 truecm
\centerline{\bf Tomohiro Matsuda
\footnote{matsuda@sit.ac.jp}}
\vskip .4 truecm
\centerline {\it Laboratory of Physics, Saitama Institute of
 Technology,}
\centerline {\it Fusaiji, Okabe-machi, Saitama 369-0293, 
Japan}
\makeatletter
\@addtoreset{equation}{section}
\def\theequation{\thesection.\arabic{equation}}
\makeatother
\vskip 1.0 truecm

\begin{abstract}
\hspace*{\parindent}
We present a new scenario for the baryon number violation that may take
place in models with large extra dimensions.
Our idea is interesting because leptogenesis with a low
reheating temperature requires an alternative source of the $B+L$
 violation to convert the produced leptons into baryons.
If one considers a model with an intermediate compactification radius,
the reheating temperature may be allowed to become higher than the critical
temperature of the conventional electroweak phase transition.
Our mechanism is still important in this case, because it provides 
 the phase boundary, which is an
essential ingredient of the electroweak baryogenesis. 
\end{abstract}

\newpage
\section{Introduction}
\hspace*{\parindent}
In spite of the great success in the quantum field theory, there is still no
consistent scenario in which the quantum gravity is included.
The most promising scenario in this direction would be the string
theory, where the consistency is ensured by the requirement of
additional dimensions.
Initially the sizes of extra dimensions had been assumed to be as small as
$M_p^{-1}$,
however it has been observed later that there is no reason
to believe such a tiny compactification radius\cite{Extra_1}.
The large extra dimension may solve the hierarchy problem.
Denoting the volume of the $n$-dimensional compact space by $V_n$,
the observed Planck mass is obtained by the relation $M_p^2=M^{n+2}_{*}V_n$,
where $M_{*}$ denotes the fundamental scale of gravity.
If one assumes more than two extra dimensions, $M_{*}$ may be
close to the electroweak scale without conflicting any observable
bounds.
Although such a low fundamental scale considerably improves the
situations of the traditional hierarchy problem, the scenario requires
some degrees of fine-tuning.
The largeness of the quantity $V_n$ is perhaps the most obvious example
of such fine-tuning.

In theories with large extra dimensions,
measured fermion masses represent a window into ultraviolet physics.
One important avenue would be to
find models in which the hierarchical patterns of the fermion masses
can be produced within natural couplings of order unity in the underlying
theory. 
There is an interesting approach to this problem, which utilizes the
locality rather than the symmetry to generate the hierarchy in the
Yukawas\cite{orbifold}.

There is an another problem related to the fine-tuning of the coupling
constant.
The problem of the $\mu$-term is important when one considers 
supersymmetric extension.

In this paper we show that some of the mechanisms, which had been put
forward to solve the above problems, can trigger the electroweak symmetry
restoration at a low temperature to activate sphaleron interactions in a
false-vacuum domain. 
We stress that the false-vacuum domain is a natural product of the
required system. 
Our mechanism is important, because it provides a source of the baryon
number violation at a low temperature.
The observed baryon number asymmetry of the Universe
requires baryon number violating interactions to have been effective but
non-equilibrium at the early stages of the Universe.
The production of the net baryon asymmetry requires baryon number
violating interactions, C and CP violation and a departure from thermal 
equilibrium\cite{sakharov}.
When the fundamental mass scale is sufficiently high, 
the first two of these ingredients are naturally contained in
conventional GUTs or other string-motivated scenarios, and the 
third can be realized in an expanding universe\cite{original}.
On the other hand, recent observations of neutrino mixing and the
measured values for the differences in mass-squareds make it more
plausible for us to include heavy Majorana neutrinos to
the Standard Model. 
These additional neutrinos can naturally be heavy since they are
singlets of the Standard Model gauge groups and their masses are not
determined by the electroweak scale.
If these heavy Majorana neutrinos had existed in the early Universe and
had effective CP violation in their decay modes, they can be natural
candidates for producing lepton asymmetry via out-of-equilibrium decays
at later period.
The leptonic asymmetry produced by the decay of these heavy Majorana
neutrinos is expected to be converted into the baryon asymmetry of the
Universe by sphaleron interactions\cite{lepto}. 
In models with supersymmetry, lepton number may be produced by the
Affleck-Dine mechanism\cite{AD}.

In models with large extra dimensions, however, the situation is
rather involved because of the requirement for the low reheating
temperature\cite{baryo_extra,lepto_extra}.
In this respect, it is very important to propose ideas for
realizing the baryon number violation at low
temperature.\footnote{Some examples are already discussed in 
ref.\cite{matsuda_ADS} for models with warped extra dimension.
Aspects of baryogenesis 
with large extra dimensions are already discussed by many authors.
For example, in ref.\cite{baryo_extra}, it is argued
that the Affleck-Dine mechanism can generate adequate baryogenesis.
In ref.\cite{dvali}, the global-charge non-conservation due to quantum
fluctuations of the brane surface is discussed. 
The baryogenesis by the decay of heavy X particle is discussed in
ref.\cite{thermalB, baryo_matsuda_defect}. 
In ref.\cite{four-point}, it is argued that a dimension-6 proton decay
operator, suppressed today by the mechanism of quark-lepton separation
in extra dimensions can generate baryon number if one assumes that this
operator was unsuppressed in the early Universe due to a time-dependent
quark-lepton separation.}  
In this paper we consider supersymmetric models with large or
intermediate extra dimensions and present a mechanism where the required
baryon number violation appears in the domain of the false vacuum where
sphalerons are activated at the temperature below
$T_{EW}$\cite{lepto_extra}.
Our idea is very simple and naturally contained in the corresponding
models for the Yukawa hierarchy\cite{orbifold} or the 
$\mu$-problem\cite{shine_mu1}.

The plan of our paper is the following.
In section 2 we show the basic idea\cite{baryo_matsuda_defect}.
In section 3 we consider the mechanism in ref.\cite{orbifold} where the 
observed fermion masses and mixings are generated by localizing the
three generations of matter and the two Higgs fields at different
locations in a compact extra dimension.
\footnote{
See also ref.\cite{Yukawa_local}}
The required domain configuration is a natural ingredient of the
model.
We show that the Yukawa hierarchy is destabilized 
and the top Yukawa is exponentially suppressed in the domain of the
degenerated false vacuum.  
Then the electroweak symmetry restoration will take place because
the radiative correction from the top is suppressed in the
domain. 
In section 4 we consider an another example where the $\mu$-problem is
solved by the localization of wavefunctions in the extra dimension.
In this case, the hierarchical tiny coupling is not
maintained in the false vacuum and large $\mu$-term appears.
Then the Higgs potential is stabilized and the electroweak
symmetry restoration takes place in the false-vacuum domain. 
\section{Localized wavefunctions and the false vacuum}
\hspace*{\parindent}
Our idea is based on ref.\cite{baryo_matsuda_defect}. 
In ref.\cite{baryo_matsuda_defect}, we argued that the localized
wavefunctions in extra dimensions are displaced in the false vacuum in
the four-dimensional spacetime so
that the exponentially suppressed interactions are enhanced in the
domain of the false vacuum.
Localization of a superfield is already discussed in 
ref.\cite{orbifold, not_orbifold_but_susy}.
\subsection{Localized wavefunctions along the fat domain wall}
\hspace*{\parindent}
To show the elements of our idea, here we limit ourselves to the 
non-supersymmetric construction with fermions localized within only one
extra dimension\cite{proton}.
To localize fields in an extra dimension, it is necessary to 
break higher dimensional translation invariance, which is
accomplished by a spatially varying expectation value of the
five-dimensional scalar field 
$\phi_{A}$ that forms a thick wall along the extra dimension.
If the scalar field $\phi_{A}$ couples to a five-dimensional 
fermionic field $\psi_{i}$ through the five-dimensional Yukawa interaction
$g \phi_A \overline{\psi}_i \psi_i$, it is possible to show that the
fermionic field localizes at the place 
where the total mass in the five-dimensional theory vanishes.
For definiteness, we consider the Lagrangian
\begin{eqnarray}
{\cal L} &=&\overline{\psi_{i}}\left(i \br{\partial_5} +g_{i}\phi_{A}(y) 
+m_{5,i}
\right)\psi_{i}\nonumber\\
&&+\frac{1}{2}\partial_{\nu}\phi_{A} \partial^{\nu}\phi_{A} \nonumber\\
&&-V(\phi_{A}),
\end{eqnarray}
where $y$ is the fifth coordinate of the extra dimension.
For the special choice $\phi_{A}(y)=2\mu^2 y$, which corresponds to 
approximating the kink with a straight line interpolating two vacua,
the wave function in the fifth coordinate becomes gaussian centered
around the zeros of $g_{i}\phi_{A}(y)+m_{5,i}$.
It is also shown\cite{Extra_1} that a chiral
fermionic field in the four-dimensional representation can result from
the localization mechanism.
When two fermions $\psi_p$ and $\psi_q$ have five-dimensional masses
$m_{5,p}$ and $m_{5,q}$, the corresponding localizations are
at $y_p=-\frac{m_{5,p}}{2g_p \mu^2}$ and $y_q=-\frac{m_{5,q}}{2g_q
\mu^2}$, respectively.
The shapes of the fermion wave functions along the fifth direction are 
\begin{eqnarray}
\Psi_p(y)&=&\frac{\mu^{1/2}}{(\pi/2)^{1/4}}
exp\left[-\mu^2 (y-y_p)^2\right]\nonumber\\
\Psi_q (y)&=&\frac{\mu^{1/2}}{(\pi/2)^{1/4}}
exp\left[-\mu^2 (y-y_q)^2\right].
\end{eqnarray}
The operator that contains both $\psi_p$ and $\psi_q$ is exponentially
suppressed in the effective four-dimensional theory.
For example, one may expect the following operator in the
five-dimensional theory,
\begin{equation}
{\cal O}_5\sim \int d^5 x [\phi_h \psi_p^5 \psi_q^5]
\end{equation}
where $\psi_p^5$ and $\psi_q^5$ are five-dimensional representations of
the fermionic fields. 
The corresponding four-dimensional operator is obtained by
simply replacing the five-dimensional fields by the zero-mode fields and
calculating the wave function overlaps along the fifth dimension $y$.
The result is
\begin{equation}
{\cal O}_4 \sim \epsilon \times \int d^4 x \phi_h \psi_p^4 \psi_q^4,
\end{equation}
where $p$ and $q$ denote the four-dimensional representations of the chiral
fermionic fields.
The small overlap of the fermionic wavefunctions along the fifth dimension 
suppresses the effective coupling, 
i.e., $\epsilon \sim e^{-\mu^2|y_p-y_q|^2}$. 

To construct a required false-vacuum configuration, we extend the above
model to include an another scalar field $\phi_B$ that determines the
five-dimensional mass $m_{5,i}(\phi_B)$. 
The field $\phi_B$ determines
the position of the center of the fermionic wavefunction along the fifth
dimension. 
We assume that $\phi_B$ does {\it not} make a kink
configuration along the fifth dimension, but {\it does} make a defect
configuration in the four-dimensional spacetime. 
For definiteness, we consider the Lagrangian
\begin{eqnarray}
{\cal L} &=&\overline{\psi_{i}}\left(i \br{\partial_5} +g_{i}\phi_{A}(y)
+m(\phi_{B})_{5,i}
\right)\psi_{i}\nonumber\\
&&+\frac{1}{2}\partial_{\nu}\phi_{k}\partial^{\nu}\phi_{k}\nonumber\\
&&-V(\phi_{k}),
\end{eqnarray}
where indices represent $i=p,q$ and $k=A,B$.
Here $\phi_{A}$ makes the kink configuration along the fifth dimension
while $\phi_B$ develops defect configuration in the four-dimensional
spacetime.
For example, $m_{5,i}$ are written as
$m(\phi_{B})_{5,i}=k_{i} \phi_{B}$ and the potential for $\phi_B$
takes the form; 
\begin{equation}
V_{B}=-m_B \phi_{B}^2 +\lambda_B \phi_B^4.
\end{equation}
In this case, because of the effective $Z_2$ symmetry of the
potential, the resultant defect is the cosmological domain wall that
interpolates between two degenerated vacua.
The center of the fermionic wavefunction in the fifth
dimension is shifted by the defect configuration of $\phi_B$
in the four-dimensional spacetime.
\subsection{Localized wavefunctions and the orbifold}
\hspace*{\parindent}
If the extra dimension is an orbifold, one can localize the
wavefunction at the fixed point\cite{orbifold, orbifold0}.
In this case one can find the required two degenerated vacua without
adding new field. 
One is the positive configuration for $0<y<L$, and the other is
the negative one.
If the sign is positive, the zero-mode is concentrated at $y=0$.
If it is negative, the zero mode is concentrated at $y=L$.
In general, two degenerated vacua generate the domain configuration.
If the hierarchical couplings are induced by the
above-mentioned mechanism of the orbifold, the hierarchy is destabilized in
the quasi-degenerated false vacuum.
\section{Destabilized fermion mass hierarchy}
\hspace*{\parindent}
In this section we consider a model for fermion masses, where
the Yukawa coupling hierarchy is generated due to the localization of
fields in extra spatial dimensions.
For our purpose, we consider the model in ref.\cite{orbifold}, which
differs from the Arkani-Hamed/Schmaltz model\cite{proton}.
In this model, small Yukawa couplings are due
to the position of each generation relative to the localized Higgs
fields and not due to the splittings of left and right handed
fermions.\footnote{ 
We notice that our idea is applicable to other models for the
Yukawa hierarchy, where the orbifold boundary condition is not 
responsible for the localization\cite{not_orbifold_but_susy,proton}.
If the Yukawa hierarchy is explained by splittings of the left and the
right handed fermions along the fat domain wall in the extra dimension,
the large Yukawa coupling may be exponentially 
suppressed in the false vacuum because the top quark wavefunction 
along the extra dimension can
be displaced by a defect configuration of $\phi_B$ in the
four-dimensional spacetime. 
The same idea is already discussed in ref.\cite{baryo_matsuda_defect} to
solve the difficulties in baryogenesis with large extra dimensions.}
Assuming that the Higgs vacuum expectation value is confined to one of
the orbifold fixed points, and that the fermions are localized with O(1)
width along the 
fifth dimension, one can obtain the hierarchical Yukawa couplings by
localizing only the third generation quarks at the Higgs boundary.
Other quarks are localized at the opposite side so that their Yukawa
couplings are suppressed.

For definiteness, we consider a Lagrangian for localizing fermionic
field,
\begin{eqnarray}
\label{orbifold1}
{\cal L} &=&\overline{\psi_{i}}\left(i \br{\partial_5} +g_{i}\phi(y)
\right)\psi_{i} +\frac{1}{2}\partial^{\mu}\phi \partial_{\mu}\phi\nonumber\\
&&-\frac{\lambda}{4}\left(\phi^2-v^2\right)^2,
\end{eqnarray}
where the couplings $g_i$ and $\lambda$ are real.
Applying the simplest set of boundary conditions in
ref.\cite{orbifold0} and compactifying on a $Z_2$
orbifold, one can obtain the localized fermions at the fixed
points at $y=0$ or $y=L$.
If $g_i \phi$ is positive, the zero-mode of $\psi_i$ is concentrated at
$y=0$. 
If it is negative, the zero mode is concentrated at $y=L$.
In this case one can obtain two degenerated solutions.\footnote{
The degeneracy is broken if there is an another scalar field $\phi'$
that satisfies the same boundary condition.
Although the $Z_2^{sim}$ symmetry that corresponds to the simultaneous
flips of $<\phi>$ and $<\phi'>$ will remain, the $Z_2$ symmetries
of their 
independent flips are explicitly broken if there exists a cross term in
the effective potential.
In this case $\phi'$ may be a constant which is called ``odd mass'' in
ref.\cite{orbifold}.
Even if the degeneracy is not destabilized, the domain wall collapses
when there is a bias when they are produced\cite{bias}.} 
Two (quasi) degenerated vacua generates a domain that interpolates
between them.
What we want to consider in this section is the cosmological
implications of the domain structure.
\footnote{Although it seems rather difficult to produce these defects
merely by the 
thermal effect after inflation, the nonthermal effect may create such
defects during the reheating period of inflation.
The nonthermal creation of matter and defect has raised a remarkable
interest.
In particular, the efficient production of such products during the period
of coherent oscillations of inflaton has been studied by many
authors\cite{PR}.
There is an another possibility that the defects are generated after the
first brane inflation, while the reheating temperature after the second
thermal brane inflation is kept much lower than the electroweak
scale\cite{thermalbrane}. 
The cosmological constraint on the domain wall that is produced before
thermal inflation is already discussed in ref.\cite{matsuda_wall}.}

If the hierarchy of fermion masses is due to the
above-mentioned localization mechanism\footnote{
See ref.\cite{orbifold} for more detail.} of the orbifold, the hierarchy
may be destabilized in the false vacuum.
If the wavefunction of the top is displaced toward the
opposite side of the extra dimension, its Yukawa coupling is 
exponentially suppressed in the false vacuum.
The small top Yukawa prevents the radiative breakdown of the electroweak
symmetry in supersymmetric theories. 
According to ref.\cite{vilenkin}, the false-vacuum domain may safely
survive until $T=\left(\frac{\sigma^2}{M_p^2}\right)^{1/4}$, 
where $\sigma$ is the tension of the corresponding domain wall.
This implies that the electroweak symmetry restoration may take place at
the temperature lower than $T_{EW}$ if the tension 
is smaller than $(10^7 GeV)^3$.
If the potential for the field $\phi$ is a flat potential,
the tension of the domain wall is $\sigma \sim m_{3/2}v^2$.
\section{Destabilized $\mu$-term}
\hspace*{\parindent}
To solve the $\mu$-problem within the setup of the extra dimensions, one
may use the shining mechanism\cite{shine_mu1, shine_mu2}. 
Here we consider a model in ref.\cite{shine_mu1}.
The MSSM matter and Higgs fields are assumed to live on a
(3+1)-dimensional brane embedded in one extra dimension.
Following ref.\cite{shine_mu2}, we employ four-dimensional N=1
superspace notation and treating the fifth coordinate $y$ as a label.
The action for a massive five-dimensional hypermultiplet $(\Phi,\Phi^c)$
is
\begin{eqnarray}
&&\int d^4 x dy \left(
\int d^4 \theta (\Phi^{\dagger}\Phi + \Phi^{c\dagger}\Phi^c)
+\int d^2\theta \Phi^c(m+\partial_y) \Phi\right)\nonumber\\
&+&\int d^4x dy \int d^2\theta \left(
-\delta(y) \sqrt{M_*} J \Phi^c 
-\delta(y-L) \sqrt{M_*} J^c \Phi 
+ \delta(y) \frac{\kappa}{\sqrt{M_*}} \Phi^c H_u H_d
\right).
\end{eqnarray}
The vacuum equations for the scalar fields are then 
\begin{eqnarray}
\Phi_F&=&-\delta(y-L)\sqrt{M_*}J^c + (m-\partial_y)\phi^c=0\nonumber\\
\Phi^c_F&=&-\delta(y)\sqrt{M_*}J + (m-\partial_y)\phi=0
\end{eqnarray}
The source $J^c$ at $y=0$ shines an expectation value for
$\phi^c$\cite{shine_mu1}, which generates the exponentially suppressed 
$\mu$-term on the matter brane.
We are interested in the case where the centers of the localized
 matter fields in the extra dimension are displaced by a cosmological
 defect in the four-dimensional spacetime.
If the center of the source $J$, or Higgs fields in the extra dimension
are displaced in the false-vacuum domain of the four-dimensional spacetime,
the hierarchical suppression for the $\mu$-term will be destabilized.
The resulting large $\mu$-term stabilizes the Higgs potential in the
effective four-dimensional theory.
Then the standard scenario of the electroweak symmetry breaking fails
and the symmetry restoration takes place in the local domain.
\section{Conclusions and Discussions}
\hspace*{\parindent}
Leptogenesis with large extra dimensions suffers a serious problem
because of the low reheat temperature that makes it impossible to
convert the produced leptons into baryons by the 
$B+L$ violating sphalerons.
In this paper we have presented a new scenario for sphaleron activation
at the temperature below $T_{EW}$.

If one considers a model with an intermediate fundamental mass,
the reheating
temperature may be allowed to become higher than the critical
temperature of the conventional electroweak phase transition.
Our mechanism is still important in this case, because it provides the
phase boundary, which is the 
essential ingredient of the electroweak baryogenesis. 

\section{Acknowledgment}
We wish to thank K.Shima for encouragement, and our colleagues in
Tokyo University for their kind hospitality.

\end{document}